\documentclass[%
 preprint,
 amsmath,amssymb,
 aps,
]{revtex4-1}

\usepackage{graphicx}
\usepackage{dcolumn}
\usepackage{bm}

\usepackage{tikz}
\usepackage{tikz-3dplot}
\usepackage{pgfplots}
\usepackage{wrapfig}
\usetikzlibrary{patterns}
\usetikzlibrary{decorations.markings}

\begin{document}

\title{Salecker-Wigner-Karolyhazy Gedankenexperiment\\ in light of the self-gravity}

\author{Michael~Maziashvili}
\email{maziashvili@iliauni.edu.ge}
\affiliation{School of Natural Sciences and Medicine, Ilia State University,\\ 3/5 Cholokashvili Ave., Tbilisi 0162, Georgia}


\begin{abstract}
 
 In Gedankenexperiment mentioned in the title, the imprecision in space-time measurement is related to the spreading of clock's wave-function with the passage of time required for the measurement. Special relativity puts a bound on the measurement time, it cannot be reduced arbitrarily as the signal used for the measurement cannot propagate with speed greater than that of light. In view of this reasoning, one is led to conclude that the clock should be heavy enough to slow down its wave-function from spreading with time. However, the general relativity puts an upper bound on clock's mass, since its size must remain greater then the Schwarzschild radius associated to it. This way one reaches a limit in length measurement.  However, as is discussed below, an additional insight into the question comes by taking into account self-gravitational effects. As a result, the uncertainty in length measurement is reduced to the Planck length.  

\end{abstract}

\pacs{Valid PACS appear here}
\maketitle

\section{Preface}

One of the characteristic features of quantum mechanical system is the presence of zero-point-fluctuations. That is, even in the ground state, physical quantities are characterized apart from their average values with the fluctuations, which are usually estimated by the mean square deviations. It is enough to mention that the vacuum fluctuations of the electromagnetic field is responsible for a number of well known phenomena. For instance, it stimulates a spontaneous emission of atom \cite{Weisskopf:1930au}, its another manifestation is Casimir force \cite{Casimir:1948dh} and the Lamb shift also can be explained by means of it \cite{Welton:1948zz}. In general, it is hard to estimate the rate of zero-point-fluctuations in quantum field theory, as it turns out to be a divergent quantity. Alternatively, one could try to use various Gedankenexperiments for estimating order of magnitude of the fluctuations of a given physical quantity. Such Gedankenmessungen usually account for the unavoidable disturbances caused by the interaction during the measuring process. One may recall a well known example of this sort of discussion concerning the electromagnetic field \cite{1933KDVS...12....3B, Bohr:1950zza}. In contrast to other fields, the metric that describes the gravitational field - determines at the same time the background space-time. Thus, one may consider the measurement of gravitational field by means of the motion of test particles \cite{Bronstein:2012zz, 1954RMxF....3..176A, Regge:1958wr, Peres:1960zz} (as is the case with electromagnetic field \cite{1933KDVS...12....3B, Bohr:1950zza}) or one may discuss the measurement of space-time characteristics like curvature \cite{Osborne:1949zz, 1960PhRv..120..643W, 1961JMP.....2..207W} and space-time intervals \cite{Wigner:1957ep, Salecker:1957be}. While there are no objections that the quantum fluctuations prevent one from measuring position with greater accuracy than the Planck length, $l_P \approx 10^{-33}$cm, \cite{Mead:1964zz}, there is still controversy about the rate of length fluctuations \cite{Karolyhazy:1966zz, Diosi:1989hy, Ng:1993jb, Diosi:1995tq, Adler:1999if, Ng:1999se, Baez:2002ra, Ng:2002up}. Karolyhazy supplemented the discussion of Salecker and Wigner \cite{Wigner:1957ep, Salecker:1957be} by noting that the minimum size of a clock is set by its Schwarzschild radius and found that the length $l$ cannot be measured with greater accuracy than $\delta l \gtrsim l_P^{2/3}l^{1/3}$ \cite{Karolyhazy:1966zz}. This result was criticized by devising new Gedankenexperiments \cite{Diosi:1989hy, Diosi:1995tq, Adler:1999if, Baez:2002ra} and supported again in a series of papers \cite{Ng:1993jb, Ng:1999se, Ng:2002up}. We are not going to discuss the counterexamples and their refutations but instead we shall argue that the bound $\delta l \gtrsim l_P$, which is considered by some authors to be the proper one, can readily be achieved by taking into account the effect of self-gravity in Salecker-Wigner-Karolyhazy Gedankenexperiment.

\section{Salecker, Wigner, Karolyhazy}

In order to demonstrate principal limitations on space-time measurement due to quantum and gravitational effects Salecker and Wigner proposed the following Gedankenexperiment \cite{Wigner:1957ep, Salecker:1957be}. The clocks are placed at the points the distance between which is being measured (the clock can be viewed as a spherical mirror inside which light is bouncing), and by measuring the time a light signal takes from one clock to another we estimate the distance between those points. Clock is characterized with some mass $m$ and radius $r_c$. Because of clock's size, the points are marked with the precision $\simeq r_c$. In addition clocks are subject to quantum fluctuations, $\delta p \simeq 1/r_c$, that give for fluctuation velocity: $\delta v \simeq 1/mr_c$. Thus, the total uncertainty in measuring the length $l=t$ (we use $\hbar = c =1$ system of units) takes the form \begin{equation} \delta l  \gtrsim r_c + l\delta v \simeq r_c + \frac{l}{mr_c}~. \nonumber \end{equation} Minimizing this equation with respect to $r_c$, one gets

\begin{equation}\label{optimal}r_c \simeq \sqrt{\frac{l}{m}}~,~~~~\delta l \simeq \sqrt{\frac{l}{m}}~.\end{equation} It seems that at the expense of mass we can always minimize the $\delta l$ as much as we want. But, as it was noticed by Karolyhazy, gravity brings new insight into the problem \cite{Karolyhazy:1966zz, Ng:1993jb}. Namely, the clock is characterized by the Schwarzschild radius $r_g \simeq l_P^2m$ and to avoid its gravitational collapse, the size of clock should be greater than its Schwarzschild radius

\begin{equation} l_P^2m \lesssim \sqrt{\frac{l}{m}}~. \nonumber  \end{equation} It gives an upper limit on $m$ \begin{equation}
m \lesssim l^{1/3}l_P^{-4/3}~, \nonumber 
\end{equation} and puts a lower bound on $\delta l$
\begin{equation}\label{lengthuncertainty} \delta l_{min} \simeq l^{1/3}l_P^{2/3}~.\end{equation} 

Let us note that the above discussion has been carried out without paying any attention to the self-gravitational effects. However, one has to draw attention to the fact that the optimal measurement in Salecker-Wigner-Karolyhazy Gedankenexperiment is done by a clock whose characteristics are very close to that of a black hole \cite{Barrow:1996gs}. If we bear in mind that it means the wave-function describing the clock to be shrunk to its Schwarzschild radius, we are driven to the conclusion that the gravitational attraction becomes very strong and it may drastically affect the wave-packet expansion. We discuss this matter in the next section.

\section{Suppresion of Wellenpaket expansion due to self-gravity}

In the above discussion the clock (as a whole) is treated as a free quantum mechanical object/body described by the Gau\ss sche Wellenpaket

\begin{equation} \psi(t,\,r) = \frac{e^{-r^2/4a^2}}{\left(2\pi\right)^{3/4}} \left[ r_c\left(1 +\frac{it}{2mr_c^2} \right)\right]^{-3/2} ~, \nonumber \end{equation} where \begin{equation} a^2 = r_c^2\left(1 +\frac{it}{2mr_c^2} \right) ~. \nonumber 
\end{equation} From this wave-packet one finds 

\begin{equation}\label{linear spread} \delta l(t) \, \simeq \, \sqrt{r_c^2 + \frac{t^2}{4m^2r_c^2}}  \,\gtrsim \,  r_c + \frac{t}{4mr_c} ~. \end{equation}

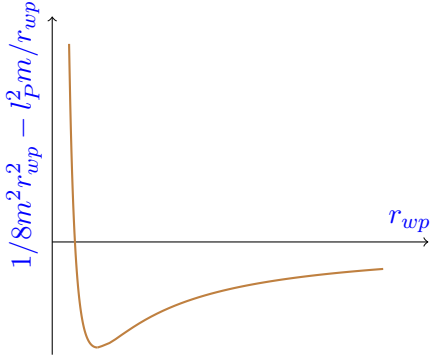
\begin{figure}[h]
\centering
\begin{tikzpicture}
      \draw[->] (0,0) -- (5,0) node[pos=0.95, blue, above] {$r_{wp}$};
      
      \draw[->] (0,-1.5) -- (0,3) node[blue, pos=0.65, sloped, above] {$1 / 8m^2 r_{wp}^2  - l_P^2m / r_{wp}$};
      
      \draw[scale=0.8,domain=0.754:5.5,smooth,variable=\x,brown, thick] plot ({\x},{1/(\x*\x) - 2.65/\x});
      
     \draw[scale=0.8,domain=0.28:0.754,smooth,variable=\x,brown, thick] plot ({\x},{1/(\x*\x) - 2.65/\x});      
    \end{tikzpicture}
\caption{The potential: $1 / 8m^2 r_{wp}^2  - l_P^2m / r_{wp}$.} \label{surati}
\end{figure}

Taking now into account the self-gravity of the Gau\ss ian wave-packet - its dynamics gets modified. For gravity prevents expansion, on general grounds one concludes that the value of $\delta l$ should be smaller than the expression \eqref{linear spread}. To get a qualitative picture, let us denote by $r_{wp}$ the radius of the wave-packet. Without gravity \begin{equation}\label{evgaussian} r_{wp}(t) \simeq \sqrt{r_c^2 + \frac{t^2}{4m^2r_c^2}}~,~~~ r_{wp}(0) = r_c~,~~~\dot{r}_{wp}(0) = 0 ~.\end{equation} The quantum mechanical acceleration responsible for this expansion has the form \begin{equation}\label{Beschleunigung} \ddot{r}_{wp}(t) = \frac{1}{4m^2\left(r_c^2 + \frac{t^2}{4m^2r_c^2} \right)^{3/2}} = \frac{1}{4m^2 r_{wp}^3}~.\end{equation} One can derive the results (\ref{optimal}, \ref{lengthuncertainty}) immediately from Eq.(\ref{evgaussian}). Minimizing the $r_{wp}(t)$ with respect to $r_c$ one gets \[ r_c = \sqrt{\frac{t}{2m}}~.\] After substituting it into Eq.(\ref{evgaussian}) one finds \[r_{wp}(t) = \sqrt{\frac{t}{m}}~. \] On the other hand, the gravitational acceleration that prevents expansion of the wave-packet looks like \cite{Carlip:2008zf} \[ a_g =   \frac{l_P^2m}{r_{wp}^2} ~.\] So that, the net acceleration takes the form 

\[a = \frac{1}{4m^2 r_{wp}^3} - \frac{l_P^2m}{r_{wp}^2} ~.\] Thus, we have to solve the equation 

\begin{eqnarray}\label{dziritadi}
\ddot{r}_{wp} =  \frac{1}{4m^2 r_{wp}^3} - \frac{l_P^2m}{r_{wp}^2} ~,~~\Rightarrow ~~ \frac{\dot{r}_{wp}^2}{2} + \frac{1}{8m^2 r_{wp}^2} - \frac{l_P^2m}{r_{wp}} = \mbox{const.}\equiv A~.\end{eqnarray} As $r_{wp}(0) = r_c,\,\dot{r}_{wp}(0) =0$, one finds 

\[ A = \frac{1}{8m^2 r_c^2} - \frac{l_P^2m}{r_c}~. \] The solution can be written in the form

\begin{equation}  \int\limits_{r_c}^{r_{wp}}\frac{dx}{\sqrt{2A + \frac{2l_P^2m}{x} - \frac{1}{4m^2 x^2}}} = t~. \end{equation} A typical form of the potential governing the dynamics of $r_{wp}$ is shown in Fig.\ref{surati}. It has a minimum at   

 \begin{equation}\label{gravopt} r_c = \frac{1}{4l_P^2m^3}~, \end{equation} corresponding to the state of stable equilibrium. Gau\ss ian wave-packet having this radius in the initial state neither contracts nor expands in course of time. From Eq.(\ref{gravopt}) one sees that the larger the mass - the smaller the clock size. However, there is an upper bound on the mass set by the Schwarzschild radius,  

\[ m_{max} \simeq \frac{r_c}{l_P^2}~, \] which together with Eq.(\ref{gravopt}) yields \[ r_c \simeq l_P~,~~~~\Rightarrow ~~~~\delta l \simeq l_P~.\]

It seems likely that one will arrive at the same result by solving the Schr\"odinger-Newton system \cite{Carlip:2008zf, Giulini:2011uw, vanMeter:2011xr}

\begin{eqnarray}\label{schrnewt} i\partial_t \psi = -\frac{1}{2m}\triangle \psi -m\varphi\psi~,~~~~\triangle \varphi = 4\pi l_P^2 m \left|\psi\right|^2 ~,\end{eqnarray} with the initial state given by the Gaussian wave-packet \[ \psi(t=0,\,r) = \frac{e^{-r^2/4r_c^2}}{\left(2\pi r_c^2\right)^{3/4}} ~.\]

It is worth noting that, apart from the above discussed effect, the self-gravity implies also the reduction of clock's mass. As this observation is significant for all discussions concerning the space-time measurements, let us confine our attention to this problem now.

\section{Reduction of mass due to self-gravity}
\label{masisdefekti}

 According to the papers \cite{Arnowitt:1962hi, Arnowitt:1960zz, Misner:1963zz, Duff:1973zz, Duff:1973ji}, we can safely say that self-gravity affects the clock's mass. The conclusion reached in the papers \cite{Arnowitt:1962hi, Arnowitt:1960zz, Misner:1963zz} implies the modification of the clock mass in the following way

\begin{eqnarray}\label{renormalized}
	m \,=\, m_c \,+\, \frac{l_P^2m^2_c}{2r_c} ~~\Rightarrow ~~ m_c \,=\, l_P^{-2}\left(\sqrt{r_c^2 +2l_P^2r_cm} \,-\, r_c\right) ~,  
\end{eqnarray} where $m$ is to be identified with the mass in absence of gravity: $l_P\to 0$. It is plain to see that $m_c$ is always positive. Duff, in his expository paper \cite{Duff:1973ji}, points out that it is not a proper conclusion and suggests the correct version in the form         

\begin{eqnarray}\label{newtonian}
m_c \,=\, m \left(1\,-\, \frac{l_P^2m}{2r_c}\right)  ~.  
\end{eqnarray} The source of this mistake is well explained in \cite{Duff:1973ji}, however, we will not dwell on the details. Instead we point out that the Eq.\eqref{newtonian} itself is very suggestive for the speculation (see \cite{Arnowitt:1962hi}) that leads to the Eq.\eqref{renormalized}. Namely, one can interpret the Eq.\eqref{newtonian} as the correction to the mass due to self-gravity in the framework of Newtonian gravity. However, one may claim that in general relativity it is the total mass that interacts gravitationally and not just the mass $m$. This way one arrives at Eq.\eqref{renormalized}. We shall consider both expressions separately.

Let us assume that the reader has no objections with regard to the Eq.\eqref{renormalized} and pose the question - how to operate with these two masses in the above discussed Gedankenexperiment? Before proceeding further, we have to make a few remarks to clarify the Eq.\eqref{renormalized}. $m_c$ is the mass that enters the exterior Schwarzschild solution. Hence, this mass determines the Schwarzschild radius. In addition, one has to require $r_c > l_P^2m/2$ in order for the solution to exist \cite{Arnowitt:1962hi, Duff:1973ji}. Thus, we demand that $r_c > l_P^2m/2$ and $r_c > 2l_P^2m_c$.

To carry the idea further, let us note that in Salecker-Wigner-Karolyhazy Gedankenexperiment the clock is described by the wave-function whose breadth is given by Eq.\eqref{evgaussian}. Therefore, $r_{wp}$ plays the role of the radius of clock-mass distribution and, accordingly, one has to replace $r_c$ in Eqs.(\ref{renormalized}, \ref{newtonian}) by this expression (recall that $r_{wp}(0)=r_c$)

\begin{eqnarray}\label{clock-mass-1}
 && m_c \,=\, l_P^{-2}\left(\sqrt{r^2_{wp} +2l_P^2r_{wp}m} \,-\, r_{wp}\right) ~, \\ \label{clock-mass-2}
 && m_c \,=\, m \left(1\,-\, \frac{l_P^2m}{2r_{wp}}\right)  ~.  
 \end{eqnarray} $m_c$ is the mass determining the gravitational field that affects the dynamics of the wave-packet. The Eq.\eqref{dziritadi} gets modified as          

\begin{eqnarray}\label{entwicklung}
 \frac{\dot{r}_{wp}^2}{2} + \frac{1}{8m^2 r_{wp}^2} - \frac{l_P^2m_c}{r_{wp}} =  \text{const}. ~. \end{eqnarray} In view of Eq.\eqref{clock-mass-1}, the one-dimensional potential governing the time evolution of $r_{wp}$ in Eq.\eqref{entwicklung} takes the form 
 
 \begin{eqnarray}
 	\frac{1}{8m^2r^2_{wp}} \,-\,  \sqrt{1 \,+\, \frac{2l_P^2m}{r_{{wp}}}} \,+\, 1  ~. \nonumber 
 \end{eqnarray} It has the same qualitative behavior as the potential depicted in Fig.\ref{surati}. It has a minimum at the point determined by the equation

 \begin{eqnarray}
 	r_c \,=\, \frac{1}{4l_P^2m^3}\sqrt{1 \,+\, \frac{2l_P^2m}{r_c}} ~. \nonumber 
 \end{eqnarray} Now $r_c$ is greater than the solution \eqref{gravopt}. From this equation it readily follows that for $m\simeq l_P^{-1}$ the minimum occurs at $r_c \simeq l_P$.

Now let us turn to the Eq.\eqref{clock-mass-2}. In this case the potential governing the dynamics of $r_{wp}$ reads

\begin{eqnarray}
	\left(\frac{1}{8m^2} \,+\, \frac{l_P^4m^2}{2}\right)\frac{1}{r^2_{wp}} \,-\, \frac{l_P^2m}{r} ~. \nonumber 
\end{eqnarray} Hence

\begin{eqnarray}
	r_c \,=\, \frac{1}{4l_P^2m^3} \,+\,l_P^2m ~, \nonumber 
\end{eqnarray} and again $r_c \simeq l_P$ for $m\simeq l_P^{-1}$.

 \section{Concluding remarks}

 The results can be summarized as follows. Salecker and Wigner found that one can always choose the size of the clock in such way that the total uncertainty in length measurement is minimized to $\delta l \simeq \sqrt{l/m}$. One can read this result also in the following way. If there is a clock of size $r_c$ and mass $m$, then the maximum distance which can be measured by this clock with accuracy $r_c$ is $r_c \simeq \sqrt{l_{max}/m}$ \cite{Barrow:1996gs} (see Eq.\eqref{optimal}). Their discussion uses the finiteness of the speed of light and the quantum mechanical expansion of the wave packet describing the clock - no mention of the effects of general relativity. Further insight into this Gedankenexperiment was obtained by Karolyhazy, who noted that minimum size of the clock is set by the Schwarzschild radius and thus one can not measure the length with greater accuracy than $\delta l \simeq l_P^{2/3}l^{1/3}$. This rate of length fluctuations is certainly much lager than $\delta l \simeq l_P$ lending thus extra interest to the issue from the standpoint of experimental signatures. It should be noted, however, that such clock is very close to the black hole and one naturally expects strong gravitational effects that will essentially affect the wave packet dynamics. We have seen that self-gravity prevents the expansion of the wave-packet and thus reduces the uncertainty in length measurement to $\delta l \simeq l_P$. One more point of importance related to self-gravity is the mass reduction. In view of the discussion presented in section \ref{masisdefekti}, we see that it does not change our conclusion made in the previous section but, in any case, it would be desirable if one could provide a numerical study of the Schr\"odinger-Newton equation by taking into account the effect of the mass reduction due to self-gravity. For this purpose one could use basic idea underlying the Schr\"odinger-Newton equation \eqref{schrnewt} as a guide. This system makes use of the Schr\"odinger equation in the background gravitational field, which in its turn is created by the mass distribution $m|\psi(t, \mathbf{r})|^2$. But the self-gravitational mass reduction implies that the gravitational field for an external observer, $r \gtrsim r_{wp}$, is sourced by the reduced mass $m_c$. Hence, one has to make the following replacement in Eq.\eqref{schrnewt}      
 
 \begin{eqnarray}
 	m|\psi(t, \mathbf{r})|^2 \, \to \, m_c|\psi(t, \mathbf{r})|^2 ~. \nonumber 
 \end{eqnarray} From Eqs.(\ref{clock-mass-1}, \ref{clock-mass-2}) it is obvious that as far as $r_{wp} \gg l_P^2m$ - the corrections are negligibly small.    
 
 In closing this section, we wanted to draw attention to the fact that the modification of Schr\"odinger-Newton system by replacing $m$ with the gravitating mass, see Eqs.(\ref{clock-mass-1}, \ref{clock-mass-2}), implies the dependence of the equation on the wave-packet breadth. The modification of Schr\"odinger equation due to quantum fluctuations of the background space suggested in \cite{Maziashvili:2016kad, Maziashvili:2018wae} is of similar nature. To stress once more our point of view, physically meaningful incorporation of $l_P$ into quantum mechanics should be expressed by some function of the ratio $l_P/r_{wp}$ rather than by a function of $l_P \langle p\rangle$, where $\langle p\rangle$ stands for average momentum. Otherwise one may obtain evidently misleading results \cite{Maziashvili:2016kad}.

\begin{acknowledgments}
Author is indebted to Avtandil Achelashvili and Zurab Kepuladze for useful discussions. 
\end{acknowledgments}

\end{document}